\documentclass[a4paper,superscriptaddress,twocolumn,showkeys,pre, aps ]{revtex4-1}

\usepackage[utf8]{inputenc}
\usepackage[english]{babel}
\usepackage{amssymb}
\usepackage{amsmath}
\usepackage{graphicx}
\usepackage{subfigure}

\usepackage[normalem]{ulem}
\usepackage{color}

\definecolor{mgray}{rgb}{0.3,0.3,0.3}

\definecolor{sigcol}{RGB}{17,97,165}
\definecolor{shadecolor}{RGB}{211,220,238}
\definecolor{edgecolor}{RGB}{17,97,165}

\newcommand{\km}{\left<k\right>}

\newcommand{\Hl}[1]{\textit{#1}}

\newcommand{\eqdot}{\,.}
\newcommand{\eqcomma}{\,,}

\newcommand{\gvec}[1]{\boldsymbol{#1}}

\renewcommand{\subsubsection}[1]{\textit{\textbf{#1.}}}

\renewcommand{\subsubsection}[1]{}

\begin{document}

\author{Kaj-Kolja Kleineberg}
\email{kkl@ffn.ub.edu}
\affiliation{Departament de F\'isica Fonamental, Universitat de Barcelona, Mart\'i i Franqu\`es 1, 08028 Barcelona, Spain}
\author{Mari\'an Bogu\~{n}\'a}
\affiliation{Departament de F\'isica Fonamental, Universitat de Barcelona, 
Mart\'i i Franqu\`es 1, 08028 Barcelona, Spain}
\date{\today}

\title{Digital Ecology: Coexistence and Domination among Interacting Networks}
\begin{abstract}
The 
overwhelming success 
of Web 2.0, within which online social networks are key actors, has induced a paradigm shift in the nature of human interactions. The user-driven character of Web 2.0 services has allowed researchers to quantify large-scale social patterns for the first time.
However, the mechanisms that determine the fate of networks at the system level are 
still poorly understood. For instance, the simultaneous existence of multiple digital services naturally raises questions concerning which conditions these services can coexist under. Analogously to the case of population dynamics, the digital world forms a complex ecosystem of interacting networks. The fitness of each network depends on its capacity to attract and maintain users' attention, which constitutes a limited resource. 
In this paper, we introduce an ecological theory of the digital world which exhibits stable coexistence of several networks as well as the dominance of an individual one, in contrast to the competitive exclusion principle. 
Interestingly, our theory also predicts that the most probable outcome is the coexistence of a moderate number of services, in agreement with empirical observations. 
\end{abstract}


\maketitle

\section{Introduction}

The rapid growth of online social networks (OSNs), such as Twitter or Facebook, has led to over two billion active accounts in 2014~\cite{digital_statshot}, and hence they can be said to cover over one quarter of the world population and $72\%$ of online U.S. adults~\cite{few_report}. 
Bridging the gap between social sciences and information and communication technologies, OSNs constitute a crucial building block in the development of innovative approaches to the challenges our current society faces.
OSNs have already changed the nature of human interactions on a worldwide scale.
In contrast to the large-scale social patterns of individuals~\cite{Bond2012,Borgatti13022009,Onnela26102010,Aral12,moro:2014}, the mechanisms underlying the fate of OSNs at the system level are not at all well understood. 

Real-world social networked systems exhibit a very high level of complexity~\cite{Barrat2008,PhysRevE.83.056109,Flammini2013,maxi2009,RevModPhys.81.591}.
In a recent study, we were able to identify the main mechanisms responsible for the evolution of quasi-isolated OSNs~\cite{our:model}. However, most OSNs operate on a worldwide scale and are in constant competition for users' attention with numerous other services; a fact that makes it extremely challenging to model them. This competitive environment leads to the extinction of some networks, while others persist. This phenomenon suggests an ecological perspective on the interaction of multiple OSNs, from which networks are considered to form a complex digital ecosystem of interacting species that compete for the same resource: users' networking time.

In standard ecology theory, Gause's law of competitive exclusion~\cite{Gause:1960} states 
that under constant environmental conditions, two species in competition for the 
same resource cannot coexist. This is because even the slightest advantage of one species over the others is amplified and eventually leads to the domination of this species. This mechanism is often referred to as rich-get-richer.
Competitive exclusion is predicted by many 
theoretical models~\cite{haken:synergetics}.
However, many observations of natural ecosystems 
seem to contradict Gause's law, as in the case of the famous plankton 
paradox~\cite{plankton:paradox}. Attempts to solve such paradoxes include the 
assumption of different roles (competition--colonization trade 
off~\cite{hastings,cadotte}), the increase of the dimension of the systems,  
the inclusion of further species properties, etc. (see~\cite{palmer} and references 
within). However, such models allow for an unlimited number of coexisting species, which thereby creates a new paradox. 
Indeed, real ecosystems usually consist of a moderate number of coexisting species. Here, we show that the coexistence of networks that are in competition for the same resource, namely our society's networking time, 
is possible. Furthermore, our work predicts that the most probable outcome is the coexistence of a moderate number of networks.

Recent work~\cite{Ribeiro:2014} showed that the 
competition between Facebook and its competitors such as MySpace in the mid 2000s 
led to the extinction of Facebook's competitors and its own prevalence. However, the current existence of a large number of OSNs~\cite{wiki_osns} suggests 
that the coexistence of multiple networks is indeed possible. This could be 
explained by analogy with the competition--colonization trade-off mentioned earlier, if we 
assume that different networks compete for different peer groups and hence 
one network can persist in each of these groups. Although the existence of 
different peer groups is certainly the case in reality, our aim in this paper 
is to introduce a 
general and concise theory
for competition between identical networks that are in competition for the same set of potential users that allows either the coexistence of any number of networks or the domination of a single network. 

We show that the coexistence of competing networks can indeed be 
modeled by allowing for the interplay of two very common mechanisms: preferential 
attachment and diminishing returns. Preferential attachment~\cite{Barabasi:1999ha,DoMeSa01,BiBa01a,CaCaDeMu02,Vazquez2003,PaSaSo03,FoFl06,SoBo07,boguna:popularity} is a fundamental 
principle that can be applied to growing networks and which states that newborn nodes are most likely 
to connect to the more popular nodes; this leads to a rich-get-richer effect.
The principle of diminishing returns---or diminishing marginal returns---is widely used in 
economic theories and refers to the negative curvature of production functions. For example, suppose that sowing 1 kilogram of seed in a certain place yields a crop of one ton. However, 2 kilograms of seed produces only 1.5 tons of crop; and 3 kilograms of seed produces 1.75 tons of crop. Thus, the marginal return per increment of seed diminishes with the increasing amount of seed used. 

In this paper, we demonstrate the following three points. First, multiple networks can coexist in a certain parameter region due to the interplay of a rich-get-richer mechanism and diminishing returns in the dynamics of the evolution of the networks. Second, we are most likely to observe only a moderate number of coexisting services. Finally, third, the influence of the mass media controls 
the observed diversity in the digital ecosystem.

\section{Results}
\subsection{From quasi-isolated online social networks to interacting networks}
\label{sec_generalization}

\begin{figure}[t]
\centering
 \includegraphics[width=1\linewidth]{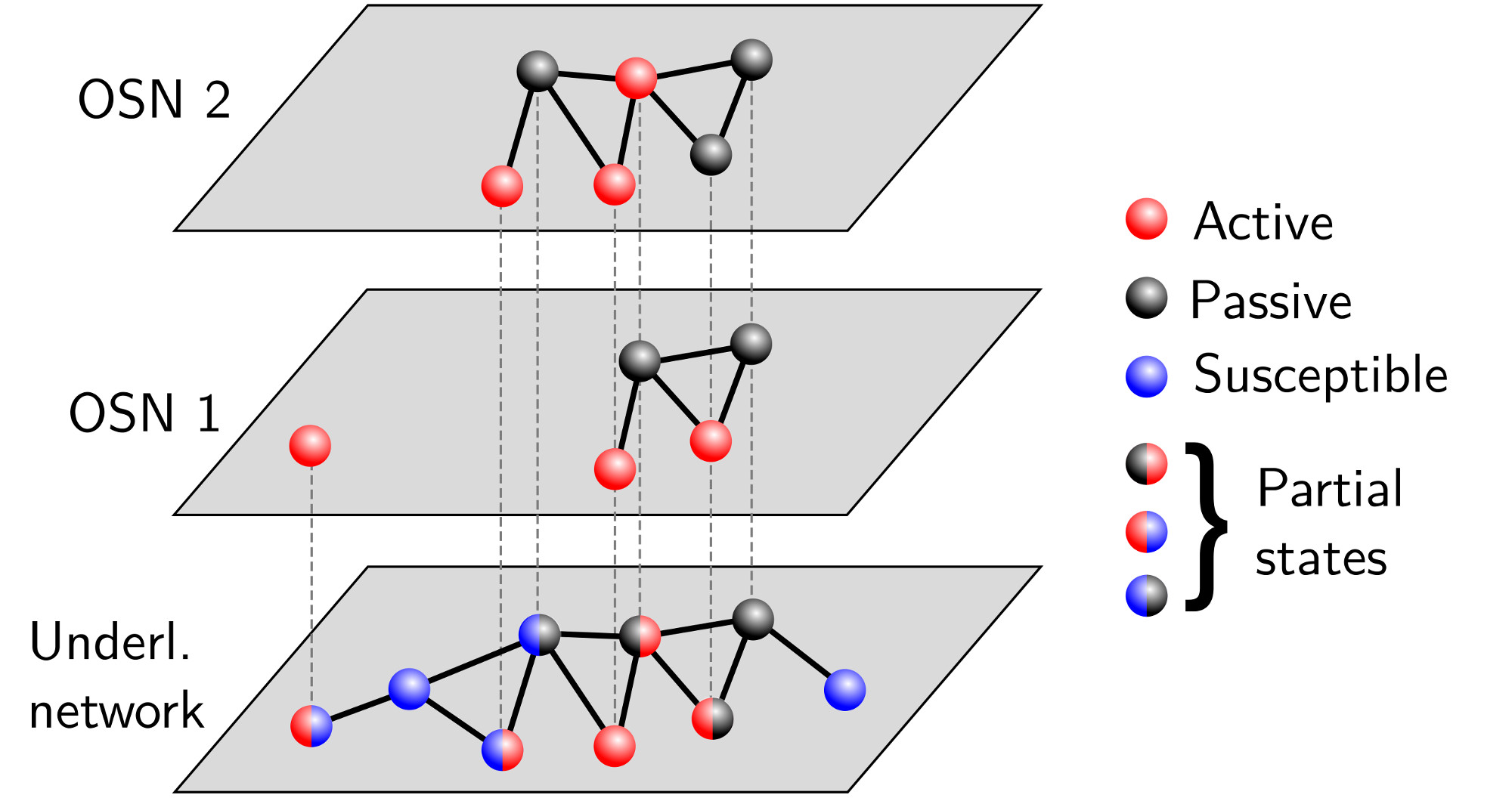}
 \caption{Multiplex layout of two online social network layers. The bottom 
layer 
represents the underlying social structure and the remaining layers represent 
each OSN. \label{fig_sketch}}
\end{figure}

\subsubsection{Evolution of quasi-isolated online social networks}
\label{sec_isolated}

The fate of a single network within the digital ecosystem depends crucially on the form of the interactions between it and its competitors, and the fitness of each of them. Nevertheless, without precise knowledge of  the evolution of a single network in the absence of competitors, little insight can be gained into the fundamental interaction mechanisms. A theory of interacting networks must therefore be built on
such precise understanding of the evolution of individual networks in isolation. In a recent study~\cite{our:model}, we were able to gauge precisely the fundamental mechanisms driving the evolution of isolated networks, which we briefly summarize in what follows.

The evolution of an OSN is coupled to the underlying social structure. The following four dynamical processes drive the evolution of the system:
\begin{enumerate}
 \item \Hl{Viral activation:} a susceptible node can be virally activated and added to the OSN by contact with an active neighbor in the traditional off-line network. Such events happen at rate $\lambda$ for each active link.
 \item \Hl{Mass media effect:} each susceptible individual becomes active spontaneously at rate $\mu$ 
and may thus be added to the OSN in response to the visibility of the OSN.
 \item \Hl{Deactivation:} active users become spontaneously passive at rate $\delta$ and no longer trigger viral activations or reactivate other passive nodes.
 \item \Hl{Viral reactivation:} at rate $\lambda$, active users can reactivate their passive neighbors. The neighbor then becomes active and can trigger both viral activations and viral reactivations. 
\end{enumerate}
The balance between the mass media influence, $\mu$, and the viral effect, $\lambda$, can be estimated from the topological 
evolution of the corresponding empirical network. This estimation can be performed by making use of the network exhibiting a dynamical percolation transition. 
The critical point of the transition depends on the ratio between $\lambda$ and $\mu$. This is due to the complementary roles the respective effects play in the topological evolution. Matching the system size at the critical point then yields a linear relationship between $\lambda$ and $\mu$ (see~\cite{our:model} for further details).

The macroscopic state of the system is characterized by the density of active nodes, defined as the quotient of active nodes over the total number of nodes, $\rho^a(t)$, and the density of passive nodes, $\rho^p(t)$. The density of susceptible nodes can be evaluated as: $\rho^s(t)=1-\rho^a(t)-\rho^p(t)$. For a detailed discussion of the model we refer the reader to Ref.~\cite{our:model}. In the present context, we want to emphasize that the model exhibits a threshold $\lambda_c$ below which the entire network eventually becomes passive.

Suppose now that, instead of a single network, there are $n_l$ networks competing for the 
same set of potential users. Each user can be active or passive in several networks simultaneously, as represented in Fig.~\ref{fig_sketch}, such that the long-term evolution of the fraction of active users in each layer determines the fate of the system: either several networks coexist or only a single network prevails. 
The first key point in the generalization of the model introduced in~\cite{our:model} concerns the role of the viral parameter $\lambda$. 
This parameter is a proxy for users' engagement in online activities, such as inviting their friends to participate in the network, generating or forwarding content, etc. However, such activities require users to spend a given amount of their time on them and their time is, obviously, bounded. This implies that when users are simultaneously active in two or more different networks or services, they are forced to decide the amount of time they devote to each of them. We model this effect by assuming that the viral parameter for each layer is $\lambda_i=\lambda \omega_i $, where $\omega_i$ a set of normalized weights (that is, $\sum_{i=1}^{n_l} \omega_i=1$) that quantify users' engagement with each OSN. In this way, $\sum_{i=1}^{n_l}\lambda_i=\lambda$ is a conserved quantity related to the physical and cognitive limitations of users. The second key point in our generalization concerns the dependence of
the share, $\lambda_i$, of the total amount of virality for individual networks on the state of activity of the whole system, which is
defined by the vector: $\gvec{\rho}^a=(\rho_1^a,\rho_2^a,\cdots,\rho_{n_l}^a)^\mathrm{T}$. We assume that the weights $\omega_i$ are functions of $\gvec{\rho}^a$ that obey the following two conditions:
\begin{enumerate}
 \item \Hl{Symmetry:} All networks are considered intrinsically equal. Therefore, the weight functions must satisfy the symmetry conditions:
\[
\begin{split}
\omega_i(\cdot,\rho_i^a,\cdots,\rho_j^a,\cdot)=\omega_j(\cdot,\rho_j^a,\cdots,\rho_i^a,\cdot)\\
\omega_i(\cdot,\rho_j^a,\cdots,\rho_k^a,\cdot)=\omega_i(\cdot,\rho_k^a,\cdots,\rho_j^a,\cdot),
\end{split}
\]
for any $i$, $j$, and $k$. This implies that when the fraction of active users is the same in all of them, the viral parameters $\lambda_i$ must also be equal in each network and, therefore, $\omega_i=1/n_l$ $\forall i$. 
\item \Hl{Preferential attachment:} We assume that users are in general more likely to subscribe to and participate in more active networks.  Hence, the weight of a given network $i$ must be a monotonically increasing function of $\rho^a_{i}$. Following the same line of reasoning, 
we also assume that a network with zero activity is not functional, so that $\omega_i(\rho_i^a=0)=0$. 
\end{enumerate}
Finally, consistent with observations in~\cite{our:model}, we assume a linear relation between $\mu_i$ and $\lambda_i$
\begin{equation}
 \mu_i = \frac{\lambda_i}{\nu} = \frac{\lambda \omega_i(\gvec{\rho}^a)}{\nu},
 \label{eqn_parameter_line}
\end{equation}
where $\nu$ denotes the relative strength of the viral effect with respect to the mass media effect (in~\cite{our:model}, we found $\nu \approx 4 \sim 5$).

These conditions can be interpreted as coarse-grained preferential attachment in the bipartite graph consisting of users and networks. Users are in general more prone to connect to networks which exhibit higher activity and, once active in more than one network, they are also more inclined to engage with the most active one more often. Notice that we are introducing a feedback loop between the global dynamics of the system and the microscopic parameters $\lambda_i$. We are thus assuming that users are, somehow, able to sense the global activity of the system. This can be achieved in practice as a combination of the amounts of information that users receive from: the network itself~\cite{weng:competition,gleeson2014,Gleeson2014pnas}, global media, the traditional off-line social network, etc. 
Although preferential attachment induces a rich-get-richer mechanism, in what follows we show  that
the interplay of this mechanism with the dynamics of the networks leads to the emergence of
stable coexistence of multiple networks across a certain parameter region.

\subsection{Mean-field approximation}
\label{sec_meanfield}

The effects of complex topologies on epidemic-like spreading processes are well understood nowadays and cannot be ignored. However, the dynamics of our model is rich and complex enough on its own to be analyzed in isolation. Therefore, in this section we perform a mean-field analysis which provides important insight into the emergence and stability of a state of coexistence of multiple networks. In particular, we replace the real social contact network by a fully mixed population with an average number of contacts per user $\langle k \rangle$. Section \ref{sec_simulations} contains numerical simulations of our dynamics using a real social network~\cite{Takac2012,our:model}. We can confirm in advance 
that the general picture drawn in this section is also observed in the real system.

\subsubsection{One-dimensional dynamics}
For one network, the system  is described by the following mean-field equations~\cite{our:model}
\begin{align}
\begin{split}
 \dot{\rho}^a & = \underbrace{\lambda \km \rho^s \rho^a}_{\text{Viral activations}} + \underbrace{\lambda \km \rho^a \rho^p}_{\text{Reactivations}} + \underbrace{\mu \rho^s}_{\text{Mass media}} - \underbrace{\delta \rho^a}_{\text{Deactivations}} \\
 \dot{\rho}^p & = -\underbrace{\lambda \km \rho^p \rho^a}_{\text{Reactivations}} + \underbrace{\delta \rho^a}_{\text{Deactivations}} \\
 \dot{\rho}^s & = \underbrace{-\mu \rho^s}_{\text{Mass media}} - 
\underbrace{\lambda \km \rho^s \rho^a}_{\text{Viral activations}}.
 \end{split}
\end{align}
The nontrivial steady-state solution is $\rho^s=0$ and $\rho^a=1-\delta/\lambda \langle k \rangle$, which is stable only when $\lambda \ge \delta/\langle k \rangle \equiv \lambda_c^1$. This defines the critical value of $\lambda$ below which activity is not possible, even in a single network. In the rest of the paper, we assume that $\lambda > \lambda_c^1$ so that, even if coexistence is not possible, at least one network is always able to survive. Likewise, we also fix the timescale of our model by setting $\delta = 1$ from now on. 

\subsubsection{Multiple competing networks}
\label{sec_coupling}

\begin{figure*}[t]
\centering
 \includegraphics[width=1\linewidth]{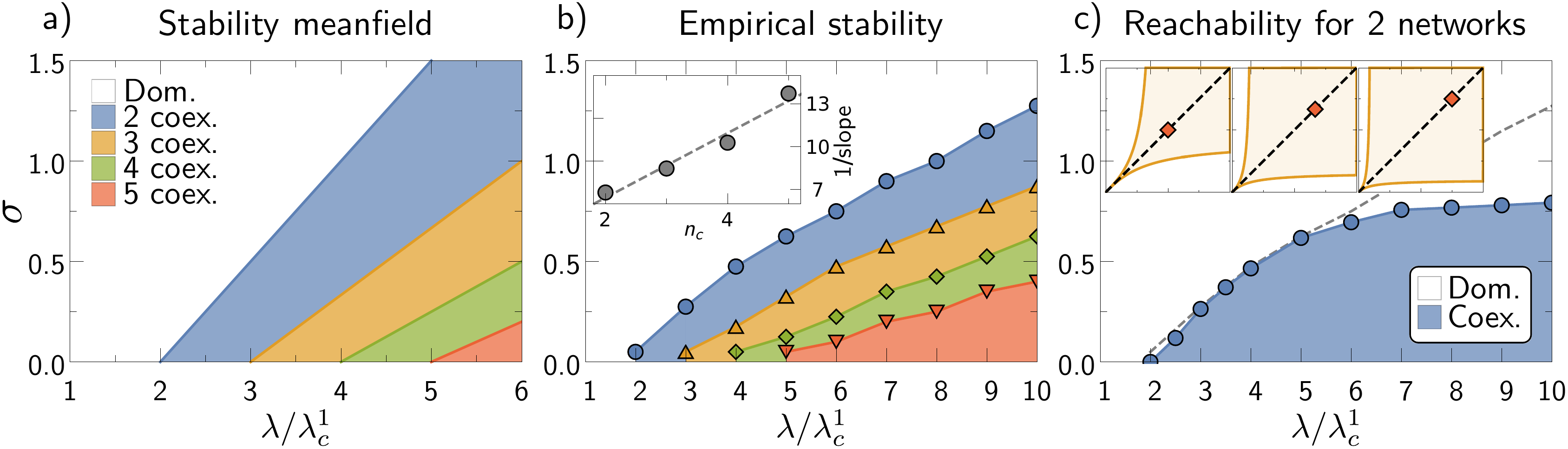}
 \caption{\textbf{a)} Regions of maximal possible coexistence in the mean-field approximation as a function of $\lambda$ and $\sigma$ for 5 networks evaluated from Eq.~\eqref{condition4stability}. \label{fig_meanfield_stability}
 \textbf{b)} Stability regions for the full stochastic model with a real underlying topology. For details see Appendix~\ref{sec_appendix_empirical_stab}. The inset shows $n_c$ versus the inverse slope of linear fits to the respective lines. \label{fig_stability_two_networks} \label{fig_empiric_stability}
 \textbf{c)} The most probable configuration reached from empty initial conditions for two networks. The dashed line corresponds to the empirical stability of the two networks. The insets ($x$ and $y$ axes each denote the activity from $0$ to $1$) show the basins of attraction in the mean-field approximation for $\sigma = 0.8$ and $\lambda/\lambda_c^1 = 4$ (left), $\lambda/\lambda_c^1 = 6$ (center), and $\lambda/\lambda_c^1 = 8$ (right).
\label{fig_4reg} \label{fig_reachability} \label{fig_attractors}}
\end{figure*}

In the case of an arbitrary number of OSNs, the system is characterized by the fraction of active and passive users in each layer, $\rho_i^a$ and $\rho_i^p$, and the fraction of individuals in the  traditional off-line social network that are susceptible to subscription in network $i$: $\rho_i^s$. We assume that the densities of active/passive/susceptible nodes are not correlated between different OSNs. Thus, the evolution equations 
in the mean-field approximation for the $i$-th layer are
\begin{align}
 \begin{split}
 \dot{\rho}^a_{i} & = \rho^a_{i} \biggr\{ \lambda \km \omega_i(\gvec{\rho}^a)  \left[ 1 - 
\rho^a_{i} \right] -1 \biggl\}+\frac{\lambda}{\nu}\omega_i(\gvec{\rho}^a) \rho_i^s\\
\dot{\rho_i^s} & = -\frac{\lambda}{\nu} \omega_i(\gvec{\rho}^a) \rho_i^s \biggr\{1+\nu \km \rho_i^a   \biggl\}
\eqcomma
\label{dynamicalsystem}
 \end{split}
\end{align}
where we have used $\rho_i^p=1-\rho^a_i-\rho_i^s$. Note that the coupling between different OSNs is encoded in the weights, $\omega_i(\gvec{\rho}^a)$. The stationary solution of Eqs.~\eqref{dynamicalsystem} that corresponds to the complete coexistence of all the $n_l$ networks is given by
\begin{equation}
\rho^{a*}_i=1-\frac{n_l}{\lambda \langle k \rangle} \; \mbox{and} \; \rho_i^{s*}=0, \; \forall i   
\label{coexistence}
\end{equation}
for $\lambda > \lambda_{c}^{n_l} \equiv \frac{n_l}{\km}$. This again defines a critical threshold for $\lambda$ below which coexistence is impossible. At the opposite extreme, the stationary solution for the prevalence of just one single network, $j$, is
\begin{equation}
\begin{split}
\rho^a_j=1-\frac{1}{\lambda \langle k \rangle} \; \mbox{and} \; \rho_j^s=0,\\
\rho_i^a=0     \; \mbox{and} \; \rho_i^s=\mbox{const} \; \; \forall i \ne j ,
\end{split}
\end{equation}
for $\lambda> \lambda_c^1$. It is easy to see that this last solution is always stable when $\lambda> \lambda_c^1$. However, the stability of the coexistence solution depends, in general, on the particular form of the weights $\omega_i(\gvec{\rho}^a)$. A detailed analysis of the Jacobian matrix of the system of Eqs.~\eqref{dynamicalsystem} evaluated at the coexistence point Eq.~\eqref{coexistence} shows that this state is stable only when
\begin{equation}
\phi(\rho_i^{a*}) \equiv \frac{n_l^2}{n_l-1}(1-\rho_i^{a*})\left. \frac{\partial \omega_i(\gvec{\rho}^a)}{\partial \rho_i^a}\right|_{\gvec{\rho}^{a*}}< 1.
\label{condition}
\end{equation}
The emergence of stable coexistence can be understood as the interplay between preferential attachment and diminishing returns. Preferential attachment affords an advantage in terms of respective weight, $\omega_i$, for networks which already exhibit higher activity; inducing a rich-get-richer effect. However, this is damped by the intrinsic dynamics of the system, which exhibits diminishing returns in terms of activity with respect to an enhancement of the corresponding weight $\omega_i$. As long as the preferential attachment mechanism is not strong enough to overcome this damping effect, any perturbation in the density of active nodes near the coexistence point will eventually decline. Hence, the coexistence is stable. 
From a mathematical point of view, this is equivalent to showing that, at the coexistence point, the function $\phi(\rho_i^{a*})$ is proportional to the dynamical return of the system when network $i$ is perturbed. In other words, if the activity of network $i$ is externally increased by a small amount $\Delta \rho_i^a$, after some relaxation time, the dynamics brings the perturbation to the new value $\Delta \tilde{\rho}_i^a = \phi(\rho_i^{a*}) \Delta \rho_i^a$. Coexistence is stable whenever the dynamical perturbation $\Delta \tilde{\rho}_i^a$ is smaller than the external one $\Delta \rho_i^a$ (see Appendix~\ref{sec_app_diminishing} for details). 
It is possible to see that $\phi(\rho_i^{a*})$ diverges at $\rho_i^{a*}=0$ and is zero when $\rho_i^{a*}=1$, and thus there is always 
a value of $\lambda$ above which
the inequality~\eqref{condition} is fulfilled (see Appendix~\ref{sec_app_existence} for details).

Interestingly, a series of states of partial coexistence exist between the complete coexistence state and the prevalence of a single network, such that only a number $n_c<n_l$ of OSNs coexist simultaneously. The symmetries of the weights $\omega_i(\gvec{\rho}^a)$ imply that any such case is exactly the same as the complete coexistence state if we replace $n_l$ by $n_c$ in Eqs.~\eqref{coexistence} and \eqref{condition}. Finally, we note that the stability of the partial or complete coexistence solutions is independent of the value of $\nu$ (see Appendix~\ref{sec_app_jacobian}). Therefore, we can discuss the stability in the limit $\nu \rightarrow \infty$, which reduces the dimensionality of the dynamical system.

The symmetry and preferential attachment conditions of the weights $\omega_i(\gvec{\rho}^a)$ combined with the normalization condition imply that, without loss of generality, $\omega_i(\gvec{\rho}^a)$ can be written as
\begin{equation}
\omega_i(\gvec{\rho}^a)=\frac{\psi(\rho_i^a)}{\sum_{j=1}^{n_l} \psi(\rho_j^a)},
\label{eqn_coupling}
\end{equation}
where $\psi$ can be any monotonically increasing function bounded on $[0,1]$ with $\psi(0)=0$. 
To gain further insight, we consider the following form of function $\psi(\rho_i^a)=\left[\rho_i^a\right]^\sigma$. 
By adjusting a single parameter this form allows us to describe a system
between a set of decoupled networks, when $\sigma=0$, and very strongly coupled ones, when $\sigma \gg 0$. In this particular case, the stability condition of the coexistence state of $n_c$ networks is given by
\begin{equation}
\sigma<\frac{\lambda-\lambda_c^{n_c}}{\lambda_c^{n_c}} \; \; \mbox{with} \;\; n_c=2,\cdots,n_l \eqdot
\label{condition4stability}
\end{equation}
This inequality defines a set of $n_l-1$ critical lines $\sigma_c(\lambda;n_c)$ in the plane $(\lambda,\sigma)$ that separate phases with $n_c$ and $n_c-1$ maximally coexisting networks. This is illustrated in Fig.~\ref{fig_meanfield_stability}a for the case of $n_l=5$ competing networks. 

\begin{figure}[t]
\centering
 \includegraphics[width=1\linewidth]{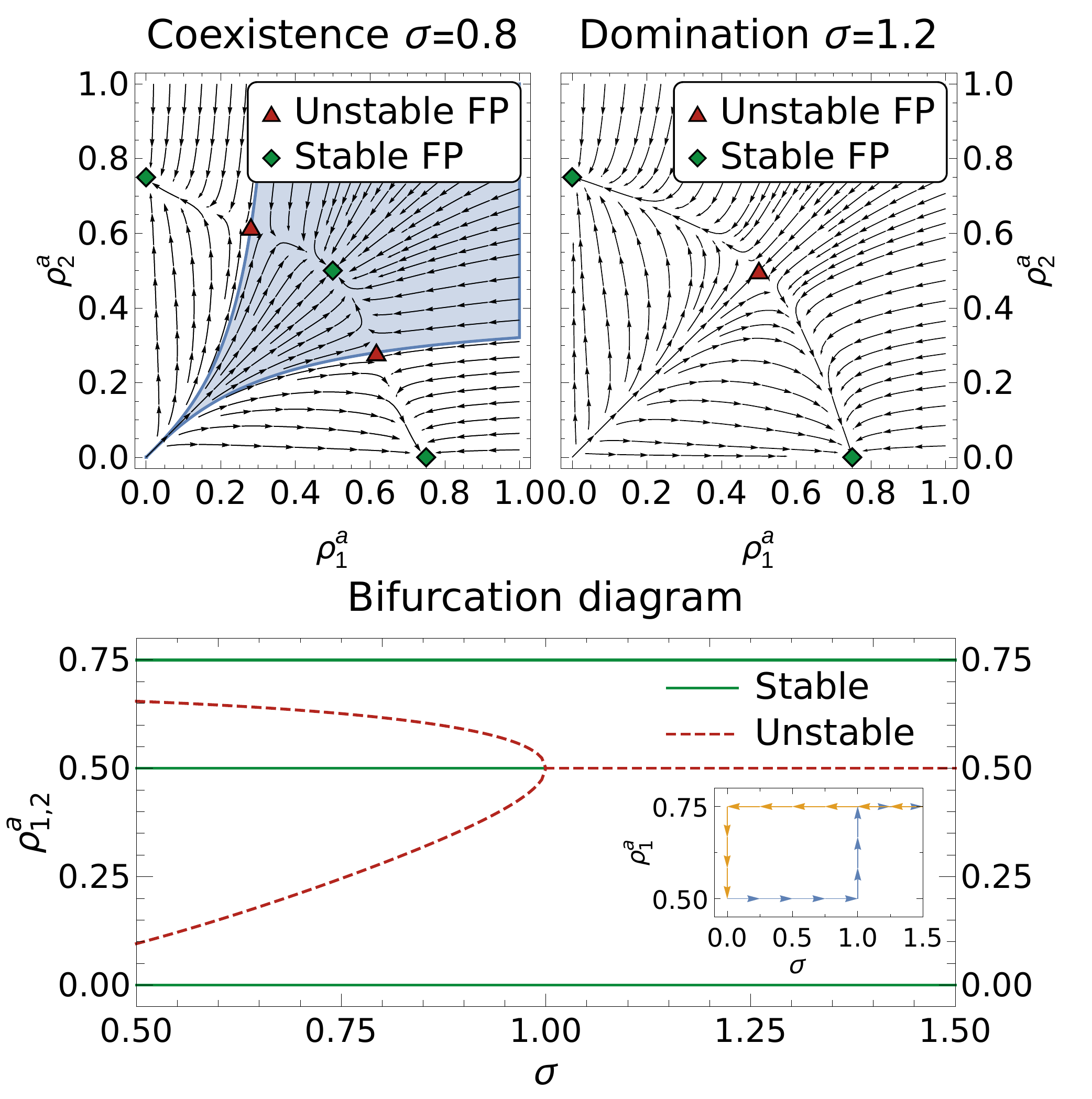}
 \caption{ Mean-field approximation in the limit $\nu \rightarrow 
\infty$ (this reduces the system dimension from 4 to 2 and allows the diagram to be plotted, see Appendix~\ref{sec_app_jacobian}). \textbf{Top:} Left: Stable 
coexistence solution ($\lambda / \lambda_c^1 = 4$, $\sigma = 0.8$). The basin of attraction for the coexistence solution is marked in blue. Right: Only the 
domination solution is stable ($\lambda / \lambda_c^1 = 4$, $\sigma = 
1.2$).
 \textbf{Bottom:} 
 Bifurcation diagram for two OSN layers showing subcritical pitchfork bifurcation at $\sigma=\sigma_c$ for $\lambda / \lambda_c^1 = 4$. The inset shows the hysteresis induced by this type of bifurcation.
\label{fig_bif}\label{fig_flow}}
\end{figure}

However, the stability of the coexistence solution does not guarantee that it is reached 
from arbitrary initial conditions because, as we show above, 
there are several other stable fixed points, each with its own basin of attraction. This is illustrated in Fig.~\ref{fig_bif}, where we show the vector field in the plane $(\rho_1^a,\rho_2^a)$ for the case of two competing networks in the limit $\nu \rightarrow \infty$. For any fixed value of $\lambda>\lambda_c^2$ and $\sigma>\sigma_c(\lambda;2)$, the coexistence solution is an unstable saddle point. This implies that one of the networks will eventually prevail, independently of the initial conditions (Fig.~\ref{fig_bif} top right). At the critical point $\sigma=\sigma_c(\lambda;2)$, the system undergoes a subcritical pitchfork bifurcation with the appearance of two unstable saddle points moving away from the (now stable) coexistence solution as $\sigma$ is decreased (Fig.~\ref{fig_bif} top left and bottom). The subcritical character of the bifurcation is akin to first-order phase transitions. Indeed, an infinitesimal increase in the value of $\sigma$ near the critical point makes the system jump from stable coexistence to the domination of one of the networks. Decreasing the value of $\sigma$ afterwards does not, however, bring the system back into the coexistence state, as this type of bifurcation implies a hysteresis effect, as shown in the inset of Fig.~\ref{fig_bif}. The two saddle points that emerge below the critical line determine the basin of attraction of the coexistence solution. This basin (depicted in blue in the top left plot of Fig.~\ref{fig_bif}) is very narrow for low densities of active nodes, as found at the beginning of the evolution. This makes the system sensitive to stochastic fluctuations; a small perturbation of the initial conditions may push the system into a state of domination of one network.

\subsection{Real-world topology}

\label{sec_simulations}

The analysis presented in the previous section is based on two strong and unrealistic assumptions: the fully mixed hypothesis of the underlying off-line social network and the absence of fluctuations in the densities of active users. The first assumption has a strong impact on the value of the critical threshold $\lambda_c^1$ and the fraction of active users in a single network when $\lambda>\lambda_c^1$. Fluctuations have an important impact mainly at the beginning of the evolution, when the number of active users is small. Such fluctuations can induce the system to change stochastically from one basin of attraction to another, leading the system to different steady states---either coexistence or domination---even if it starts from the same initial configuration with identical parameters. To understand the effects of these assumptions within a real scenario, we performed large-scale numerical simulations of our model on a real social network, the Slovakian friendship-oriented OSN Pokec~\cite{Takac2012,our:model} in 2012. The size of this network ($1.2\times10^6$ users) represents 25\% of the population of Slovakia but demographic analysis shows that it covers a much larger fraction of the population susceptible to ever participate in OSNs. This makes Pokec a very good proxy of the underlying social structure.

We first study the coexistence space in the plane $(\sigma,\lambda)$ in the case $n_l=5$. To do so, for each value of $\lambda$ and $\sigma$ we first set the system to the coexistence solution $\gvec{\rho}^{a*}$. We then apply a small positive perturbation to one of the networks $\rho_1^{a*} \rightarrow \rho_1^{a*}+\delta \rho_1^a$. The evolution of the system after this perturbation can be used to determine the stability of the coexistence state (simulation details can be found in the Appendix~\ref{sec_app_simulation_details}). The results are shown in Fig.~\ref{fig_stability_two_networks}b. Even though the position of the critical point of a single network $\lambda_c^1$ of the real Pokec network 
is extremely different from the mean-field prediction, the critical lines as a function of the ratio $\lambda/\lambda_c^1$ follow a linear trend, as in the mean-field prediction. Interestingly, the slopes of these lines (although they are different from those in the mean-field case) scale with $n_l$ in the same way as in the mean-field case (see the inset in Fig.~\ref{fig_empiric_stability}b).

However, the stability of the coexistence solution {\it per se} does not guarantee that coexistence is reached from any initial configuration. This is particularly relevant when the evolution starts from empty networks, as fluctuations in the number of active users at the beginning of the evolution can induce the system to jump from one basin of attraction to another. Therefore, to determine the effective coexistence space in the plane $(\sigma,\lambda)$, we evaluate the probability that a state of coexistence of a certain number of networks is reached when starting from empty networks. In the case of two competing networks, we define the effective critical line $\sigma_{c}^{eff}(\lambda;2)$ as the line below which the probability of the two networks reaching coexistence is greater than $1/2$.

\begin{figure}[t]
\centering
 \includegraphics[width=0.85\linewidth]{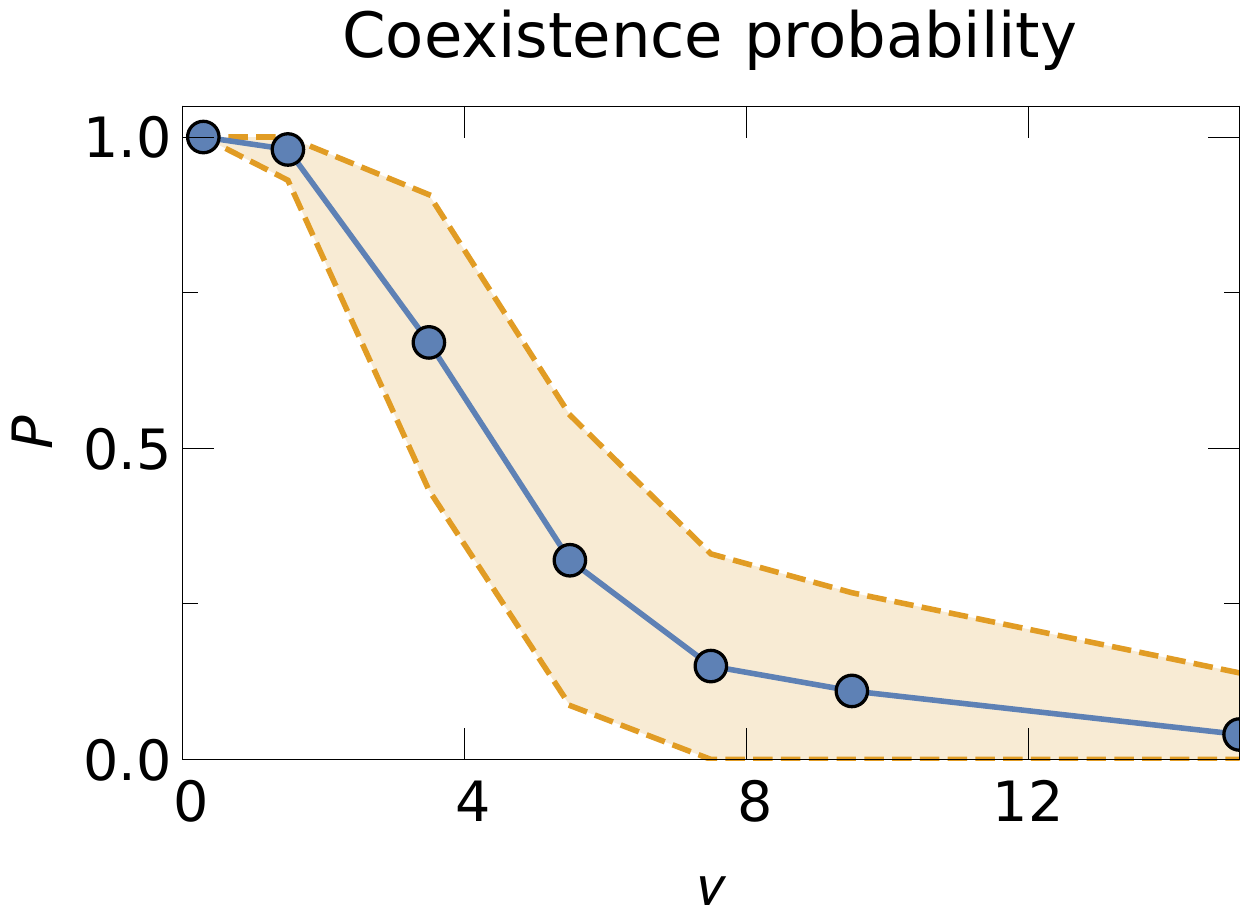}
 \caption{Probability of reaching the coexistence state for two networks for different values of $\nu$, for $\lambda / \lambda_c^1 = 6$ and $\sigma = 0.70$. The yellow area denotes one standard deviation (from top to bottom).
\label{fig_reachability_nu}}
\end{figure}

Figure~\ref{fig_4reg}c shows the results of this program for two competing networks and $\nu=4$. The effective critical line follows the critical line in Fig.~\ref{fig_stability_two_networks}b for low values of $\lambda$ and saturates at a constant value when $\lambda/\lambda_c^1 \gg1$. This result can be understood in terms of the shape of the basin of attraction of the coexistence solution near the origin. Indeed, only in this region are fluctuations important enough to make the system change from one basin to the other. As an illustration, in the inset of Fig.~\ref{fig_attractors}c  we show such a basin for $n_l=2$ and different values of $\lambda$ in the mean-field approximation. As can be observed, the shape of the basin in the neighborhood of $\rho^a_{1,2} \sim 0$ is almost independent of the value of $\lambda$, which explains why the probability of reaching the coexistence state saturates at a constant value. 

This saturation effect is similarly observed for systems of more networks, where the effective critical lines of higher coexistence states successively saturate at lower values; that is $\sigma_c^{eff}(\infty;2)>\sigma_c^{eff}(\infty;3)>\sigma_c^{eff}(\infty;4) \cdots$, which narrows the effective coexistence region in the plane $(\lambda,\sigma)$ for large numbers of networks. This is particularly relevant because, although our theory allows for the coexistence of an arbitrarily large number of networks, the stochastic nature of the dynamics, coupled with the narrow form of the basin of attraction at low densities of active users, makes such coexistence highly improbable. Therefore, our model predicts---even without knowledge of the exact empirical parameters---a moderate number of coexisting networks in a large fraction of the parameter space.

The results shown in Fig.~\ref{fig_stability_two_networks}c are obtained for a fixed value of the parameter $\nu$. While this parameter has no influence on the stability of the coexistence solution, and thus no effect on the results shown in Fig.~\ref{fig_stability_two_networks}b, it has a strong influence on the probability of reaching coexistence. Indeed, when $\nu$ is finite, the last term in Eq.~\eqref{dynamicalsystem} acts, at the beginning of the evolution, as a temporal boost that increases the fraction of active users in each network.
This mechanism drives the system closer to the coexistence state where its attractor is broader. 
Figure~\ref{fig_reachability_nu} shows the simulation results of the probability of reaching coexistence as a function of $\nu$ for two competing networks. For small values of $\nu$, the initial boost is large and the system almost always ends up in the coexistence state. For larger values of $\nu$, the probability decreases significantly. 
We conclude that a higher boost---hence a smaller value of $\nu$---favors the effective reachability of the coexistence state; whereas a small boost reduces that probability dramatically.
Since $\nu$ is related to the influence of mass media, these results show that mass media influence 
plays a crucial role in the diversity of the digital ecosystem. 

\begin{figure}[t]
 \centering
 \includegraphics[width=1\linewidth]{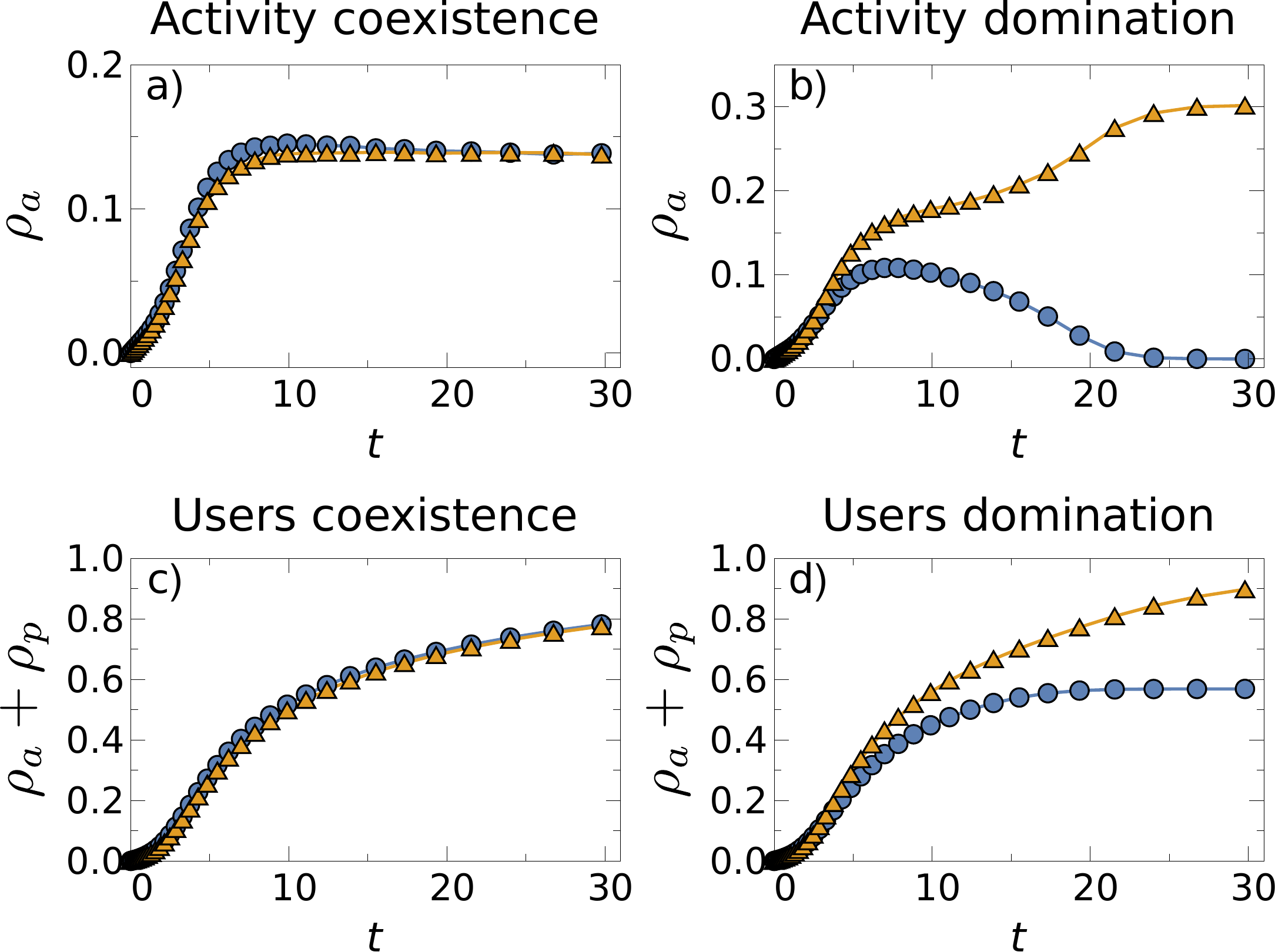}
 \caption{Evolution of the fraction of active users (top) and the fraction of total users (bottom) for two competing networks. The first column corresponds to the parameters $\lambda/\lambda_c^1 = 5$, $\sigma = 0.5$, and $\nu=4$ which lies in the coexistence region. The second column represents the parameters $\lambda/\lambda_c^1 = 5$, $\sigma = 0.75$, and $\nu=4$, which lies in the dominance region. \label{fig_2_hom_dom}}
\end{figure}
\begin{figure}[t]
 \centering
 \includegraphics[width=1\linewidth]{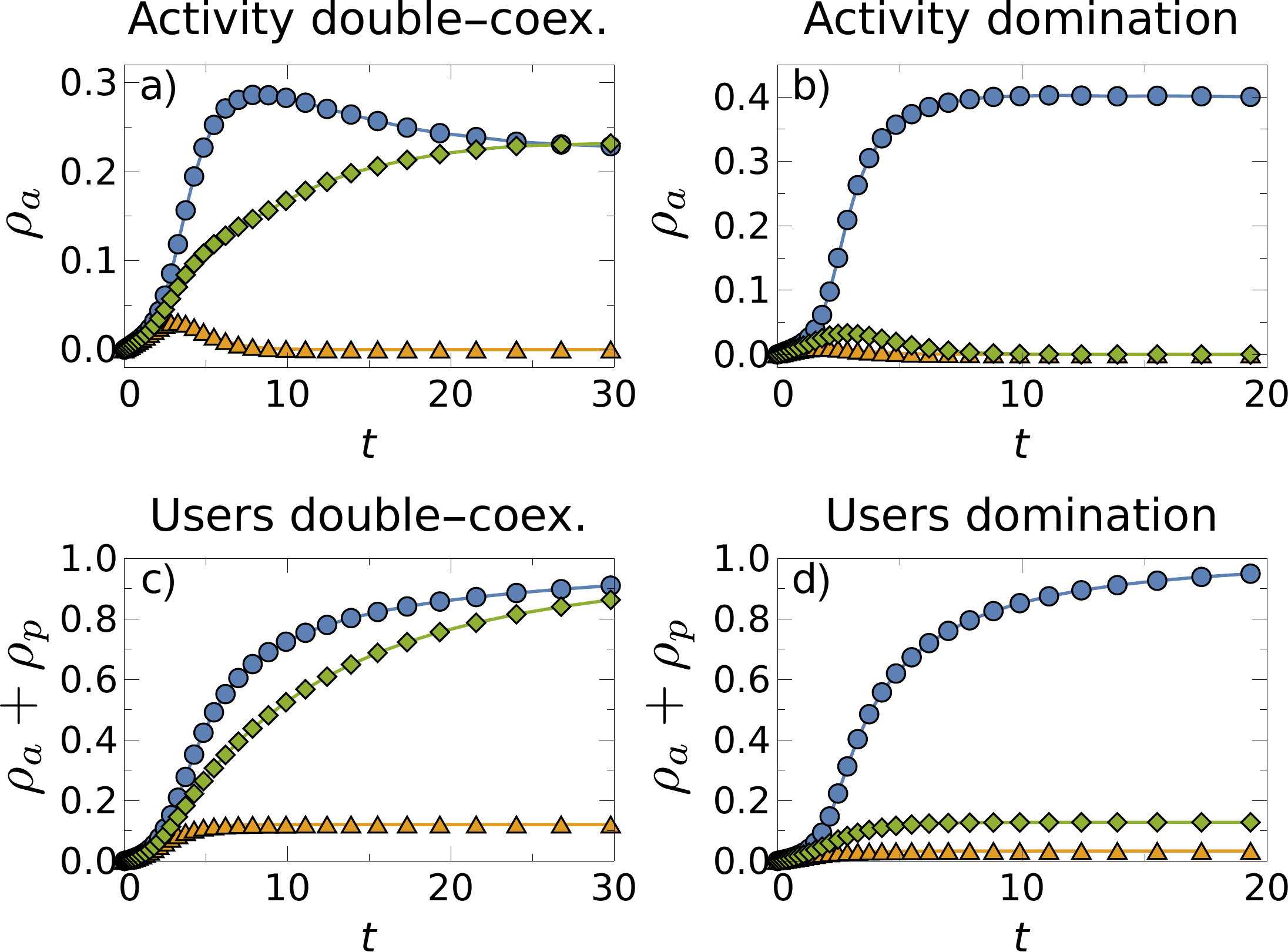}
 \caption{Evolution of the fraction of active users (top) and the fraction of total users (bottom) for three competing networks. Both columns correspond to the same parameters $\lambda/\lambda_c^1 = 7.5$,  $\sigma = 
0.8$, $\nu = 4$, but are different realizations.\label{fig_3nw_explicit}}
\end{figure}

The temporal evolution of the process also shows interesting patterns. Figure~\ref{fig_2_hom_dom} shows typical realizations 
of the process below and above the effective critical line in the case of two competing networks. It should be noted that in both cases, during the first stage of the evolution, the two networks acquire a very similar number of active users, making the forecasting of which network will eventually prevail very difficult. In a second stage, the symmetry is broken and one of the networks starts dominating, while the activity of the other declines. This pattern of ``rise and fall'' has been observed in many real OSNs~\cite{Bruno:Ribeiro}. In our model, however, such behavior is a consequence of the non-linear coupling between the networks, without the need to introduce an exogenous mechanism to explain it~\cite{Ribeiro:2014}. Meanwhile, the effective critical lines shown in Fig.~\ref{fig_4reg}c separate regions in a probabilistic way. This implies that in the vicinity of these lines, it is possible to find realizations 
that, with the same parameters and initial conditions, have opposite fates. This is illustrated in Fig.~\ref{fig_3nw_explicit} where we show two different realizations 
of three competing networks. In the first column of Fig.~\ref{fig_3nw_explicit}, we show one such realization 
where two out of three networks coexist and, in the second column, a realization 
where only one of the three networks prevails.

\section{Discussion}
\label{sec_discussion}
OSNs constantly compete to attract and retain users' attention. From this point of view, OSNs and other digital services can be understood as forming a complex digital ecosystem of interacting species that compete for the same resource: our networking time. 
In this work, we have introduced a very 
general and concise theory of
such an ecosystem. Akin to standard ecological theories of competing species, the fitness of OSNs increases with their performance following a preferential attachment (or rich-get-richer) mechanism. However, unlike the case of standard ecology, the total fitness of the system is a conserved quantity, which induces diminishing returns in the fitness of each network. Over a range of parameters, the combination of these two mechanisms leads to stable states of coexistence of many networks, in stark contrast to the competitive exclusion principle~\cite{Gause:1960}.

However, stable coexistence is only possible 
across a range of the parameter space, which is delimited by a critical line. At that critical line the system undergoes a subcritical pitchfork bifurcation akin to a first-order phase transition. Our model thus predicts that a minimal change or perturbation in the interactions between the different networks can have a catastrophic effect on the fate of the system.
In any case, due to the stochastic nature of the dynamics and the multitude of fixed points, a stable coexistence solution is not always reached. The probability of reaching such a solution is an indicator of the diversity observed in the digital ecosystem.
Interestingly, we find that over a large proportion of the parameter space the most probable outcome is the coexistence of a moderate number of digital services; in agreement with empirical observations. This number is, in general, greatly affected by the magnitude of the mass media influence.

The flexibility of our theory allows us to reproduce, with only three parameters, a large number of possible outcomes that have been observed empirically. Moreover, it can easily be modified to account for more complex situations, such as intrinsic differences between the networks or different launch times. This would allow an understanding to be gained of the extent to which higher intrinsic performance of one network can overcome the launch time advantage of another. 
Our model can also be extended to incorporate a description of the worldwide ecology of OSNs by incorporating different underlying societies that would represent different countries. It remains a task for future research to validate our assumptions regarding the coarse-grained coupling mechanism.

\begin{acknowledgments}
This work was supported by: the European Commission within the Marie Curie ITN ``iSocial'' grant no.\ PITN-GA-2012-316808; a James S. McDonnell Foundation Scholar Award in Complex Systems; the ICREA Academia prize, funded by the {\it Generalitat de Catalunya}; the MINECO projects nos.\ FIS2010-21781-C02-02 and FIS2013-47282-C2-1-P;  and the {\it Generalitat de Catalunya} grant no.\ 2014SGR608. Furthermore, M.~B. acknowledges support from the European Commission LASAGNE project no.\ 318132 (STREP).
\end{acknowledgments}

\appendix

\section{Numerical simulations}

\label{sec_app_simulation_details}
\label{sec_app_simulation_details_general}
To simulate our model we use the Pokec network~\cite{our:model}, a real OSN, as the underlying off-line network. We take advantage of the fact that the temporal events in our model are independent Poisson point processes, which allows us to use the Gillespie~\cite{Gillespie1,Gillespie2} algorithm (also known as the Doob-Gillespie~\cite{Doob1} algorithm). For a single network, the algorithm works as follows:

\begin{enumerate}
 \item Initialize the system and fix the rates corresponding to the respective events (here $\lambda$, $\mu$, $\delta = 1$)
 \item \begin{enumerate}
 \item Evaluate the number of susceptible nodes ($N_S$), the number of active nodes ($N_A$) as well as the number of edges connecting susceptible and active nodes ($E_{SA}$) and the number of edges connecting active and passive nodes ($E_{PA}$). Evaluate the sum $\mathcal{S} = \mu N_S + \lambda \left[E_{SA} + E_{PA}\right] + \delta N_A$.
 \item Generate random numbers to choose the next event. The probabilities for the events are the following: \begin{itemize}
                                                                                                            \item Mass media activation: $\mu N_S / \mathcal{S}$  
                                                                                                            \item Viral activation: $\lambda E_{SA}/ \mathcal{S}$
                                                                                                            \item Viral reactivation: $\lambda E_{PA}/ \mathcal{S}$
                                                                                                            \item Deactivation: $\delta N_A/ \mathcal{S}$
                                                                                                           \end{itemize}
\item Evaluate the corresponding time step $\tau$. The corresponding time step is given by \hbox{$\tau = \mathcal{S}^{-1}$.}                                                                                                      
\end{enumerate}
\item Update the status of the system.
So, if in step 2 a mass media activation was chosen, we randomly choose a susceptible node and change its status to active. For a deactivation, we randomly choose an active node which then becomes passive. 
In the case of viral activation, we randomly choose an edge connecting a susceptible and an active node and activate the susceptible node at the end of the link. Similarly, for viral reactivation we randomly choose an edge connecting a passive and an active node and the passive node at the end of this edge becomes active. We increase the time: $t \rightarrow t + \tau$. We iterate by returning to step 2 until the end of the simulation is reached.
 \end{enumerate}
Generalization of the algorithm to multiple layers is straightforward. One evaluates the probabilities of having a certain dynamical process in a certain layer; for example, mass media activation in layer $i$ occurs with probability $\mu_i N_{S,i}$, where $N_{S,i}$ denotes the number of susceptible nodes with respect to layer $i$ (all the nodes which are in the underlying network but not in the $i$-th layer) and $\mu_i$ is the corresponding rate in layer $i$. Accordingly, the probability of viral activation in layer $i$ is given by $\lambda_i E_{SA,i}$, where $E_{SA,i}$ is the number of edges connecting active and susceptible nodes in layer $i$. One then chooses a dynamical processes in a certain layer in accordance with these probabilities. Finally, $\tau$ is given by the inverse of the sum over all these probabilities in all the layers.

\begin{figure}[b]
\centering
 \includegraphics[width=0.85\linewidth]{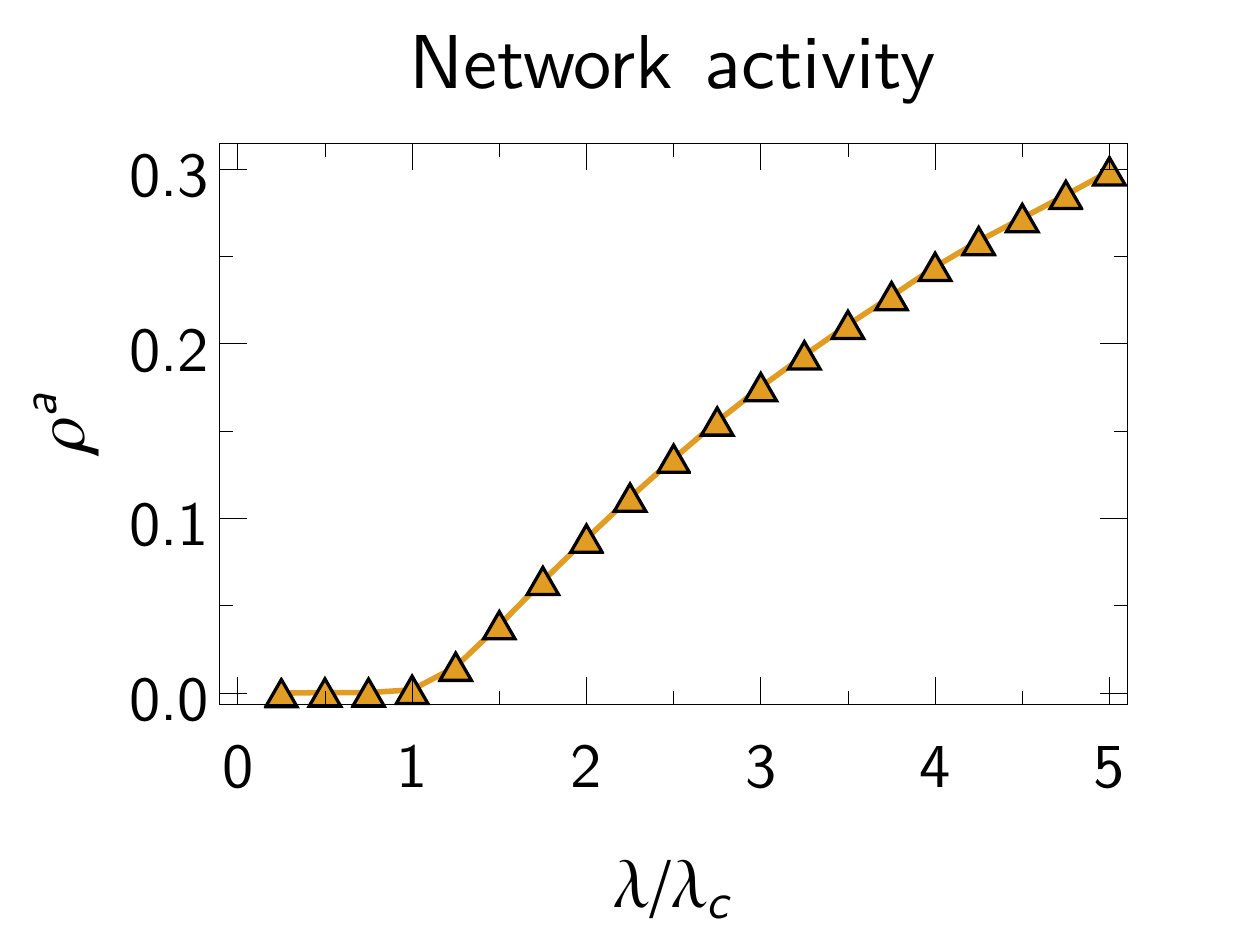}
 \caption{Steady state activity of our model for a single network when using Pokec as the underlying off-line network as a function of the virality parameter $\lambda$ (see~\cite{our:model}). Below a critical value $\lambda_c$, the network becomes 
entirely passive.  \label{fig_threshold}}
\end{figure}

\label{sec_appendix_empirical_stab}
\label{sec_app_simulation_details_empirical_stability}

Independent of the topological properties of the network, the activities for the steady state solution for an arbitrary number of layers is encoded in the activity curve of a single layer, which we show in Fig.~\ref{fig_threshold}. At the  steady state of $n_c$ coexisting networks, each prevailing layer has the same share of the total virality $\lambda_i = \omega_i \lambda = \frac{\lambda}{n_c}$, whereas the remaining ones have $\lambda_j = 0$. The steady state activity of the $i$-th network is then given by the activity value of a single layer shown in Fig.~\ref{fig_threshold} at $\lambda = \lambda_i$.

\section{Jacobian matrix}
\label{sec_app_jacobian}

Here we analyze the Jacobian matrix of the dynamical system defined in the paper whose entries correspond to the following derivatives
\begin{align}
 \begin{split}
   \left.\frac{\partial \dot{\rho}_i^a}{\partial \rho_j^a}\right|_{\gvec{\rho}^*} & = \rho_i^{a*} n_l \left.\frac{\partial \omega_i(\gvec{\rho}^a)}{\partial \rho_j^a}\right|_{\gvec{\rho}^{*}} - \delta_{ij}\frac{ \rho_i^{a*} }{1-\rho_i^{a*}} \\
 \left. \frac{\partial \dot{\rho}_i^a}{\partial \rho_j^s}\right|_{\gvec{\rho}^*} & = \delta_{ij} \frac{\lambda}{\nu n_l} \\
  \left.\frac{\partial \dot{\rho}_i^s}{\partial \rho_j^a}\right|_{\gvec{\rho}^*} & = 0 \\
   \left. \frac{\partial \dot{\rho}_i^s}{\partial \rho_j^s}\right|_{\gvec{\rho}^*} & = -\delta_{ij} \left[ \frac{\lambda}{\nu n_l} + \frac{\rho_i^{a*}}{1-\rho_i^{a*}} \right]  \eqdot
 \end{split}
\end{align}
The Jacobian matrix (of dimension $2 n_l \times 2 n_l$) can be written as
\begin{equation}
\mathcal{J} = 
 \begin{pmatrix}
  \mathcal{M}_{a,a} & \mathcal{M}_{a,s} \\
  \mathcal{M}_{s,a} & \mathcal{M}_{s,s}
 \end{pmatrix}
\end{equation}
where $\mathcal{M}_{\eta,\theta}$ represent $n_l \times n_l$ matrices with the following elements
\begin{equation}
  \mathcal{M}_{\eta,\theta}(i,j)  = \left.\frac{\partial \dot{\rho}_i^\eta}{\partial \rho_j^\theta}\right|_{\gvec{\rho}^*} \eqdot
\end{equation}
The matrix $\mathcal{M}_{s,s}$ is a diagonal matrix with all its diagonal 
equal to $d = -\left[ \frac{\lambda}{\nu n_l} + \frac{\rho_i^{a*}}{1-\rho_i^{a*}} \right] < 0$. $\mathcal{M}_{a,s}$ is also a diagonal matrix with all its diagonal elements equal to some value $c$ and finally $\mathcal{M}_{s,a}$ has all its elements equal to $0$.

By using the Laplace expansion starting from the $2n_l, 2n_l$ entry and expanding row-wise, one finds after $n_l$ iterations that
\begin{equation}
 \det\left[ \mathcal{J} - \Lambda \mathcal{I} \right] = (d-\Lambda)^{n_l} \det\left[ \mathcal{M}_{a,a}  - \Lambda \mathcal{I}  \right]
\end{equation}
which means we have the eigenvalues $\Lambda_{n_l+1,\dots,2n_l} = d$ with degeneracy $n_l$ and the remaining eigenvalues are those of $\mathcal{M}_{a,a}$.
Because $d < 0$, this means that the stability of the coexistence solution is exclusively determined by the dynamics in the limit $\nu \rightarrow \infty$, which reduces the dimensionality of the system from $2n_l$ to $n_l$ and decouples the dynamics of $\rho_i^a$ from $\rho_i^s$. See Fig.~4 in the paper for the dynamical properties.

The matrix $\mathcal{M}_{a,a}$ has the form
\begin{equation}
 \mathcal{M}_{a,a} = 
 \begin{pmatrix}
  \alpha & \beta & \cdots & \beta \\
  \beta & \alpha & \cdots & \beta \\
  \vdots  & \vdots  & \ddots & \vdots  \\
  \beta & \beta & \cdots & \alpha
 \end{pmatrix}
\end{equation}
and its eigenvalues are $\Lambda_1 = \alpha+(n_l-1) \beta$ and $\Lambda_{2,\dots,n_l} = (\alpha-\beta)$, which has degeneracy $n_l-1$.
With
\begin{equation}
 \alpha = \rho_i^{a*} n_l \left.\frac{\partial \omega_i(\gvec{\rho}^a)}{\partial \rho_i^a}\right|_{\gvec{\rho}^{*}} - \frac{\rho_i^{a*}}{1-\rho_i^{a*}}
\end{equation}
and
\begin{equation}
 \beta = \rho_i^{a*} n_l \left.\frac{\partial \omega_i(\gvec{\rho}^a)}{\partial \rho_j^a}\right|_{\gvec{\rho}^{*}} = -\rho_i^{a*}  \frac{n_l}{n_l-1} \left.\frac{\partial \omega_i(\gvec{\rho}^a)}{\partial \rho_i^a}\right|_{\gvec{\rho}^{*}}  
\end{equation}
(here we used $\sum_j \omega_j = 1$, hence $\frac{\partial \omega_i}{\partial \rho_j^a}|_{i \neq j} = -\frac{1}{n_l-1}  \frac{\partial \omega_i}{\partial \rho_i^a}$). From here, one obtains the result that $\Lambda_1$ is always negative and the stability is controlled by the eigenvalues $\Lambda_{2,\dots,n_l}$, which are
\begin{equation}
 \Lambda_{2,\dots,n_l} = \alpha - \beta  = \left.\frac{\partial \omega_i(\gvec{\rho}^a)}{\partial \rho_i^a}\right|_{\gvec{\rho}^{*}} \rho_i^{a*}  \frac{n_l^2}{n_l-1}- \frac{\rho_i^{a*}}{1-\rho_i^{a*}}
\end{equation}
which have to be negative to satisfy stable coexistence. This leads to the condition
\begin{equation}
 \frac{n_l^2}{n_l-1} (1-\rho_i^{a*}) \left.\frac{\partial \omega_i(\gvec{\rho}^a)}{\partial \rho_i^a}\right|_{\gvec{\rho}^{a*}} < 1 
\end{equation}
as described in Eq.~(6) in the paper.

\section{Dynamical response to a small perturbation for $n \gg 1$}
\label{sec_app_diminishing}

To evaluate the stability of the system in the coexistence state, one has to perform an analysis of the Jacobian matrix as shown in section~\ref{sec_app_jacobian}. In this section, we discuss the response of the system to a small perturbation, $\Delta \rho_i^a$, of the activity of one network. In the limit $n \gg 1$, we can neglect the effect of perturbing the $i$-th network on the remaining ones. The perturbation induces a shift in the corresponding weight according to
\begin{equation}
 \Delta \omega_i \approx \left.\frac{\partial \omega_i}{\partial \rho_i^a}\right|_{\gvec{\rho}^{a*}} \Delta \rho_i^a \eqdot
 \label{eqn_delta_omega}
\end{equation}
Our initial perturbation triggers the dynamical response $\Delta \tilde{\rho}$ from the system given by
\begin{equation}
 \Delta \tilde{\rho}_i^a = \left.\frac{\partial \rho_i^{a*}(\omega_i)}{\partial \omega_i}\right|_{\omega_i=\frac{1}{n_l}}  \Delta \omega_i \eqcomma
\end{equation}
where $ \rho_i^{a*}(\omega_i)=1-1/\lambda \langle k \rangle \omega_i$.
With Eq.~\eqref{eqn_delta_omega}, we obtain
\begin{equation}
 \Delta \tilde{\rho}_i^a = n_l (1 - \rho_i^{a*}) \left.\frac{\partial \omega_i}{\partial \rho_i^a}\right|_{\gvec{\rho}^{a*}} \Delta \rho_i^a. 
 \end{equation}
The coexistence solution is stable if the perturbation decreases; this means that the dynamical response $\Delta \tilde{\rho}_i^a$ has to be smaller than the initial perturbation $\Delta \rho_i^a$. Mathematically, this leads to the condition
\begin{equation}
  n_l (1-\rho_i^{a*}) \left.\frac{\partial \omega_i}{\partial \rho_i^a}\right|_{\gvec{\rho}^{a*}} < 1
 \label{eqn_reduced_condition}
\end{equation}
which is equivalent to Eq.~(6) in the paper in the limit $n_l \gg 1$. The left-hand side of Eq.~\eqref{eqn_reduced_condition} is proportional the ratio between the dynamical response of the system and the initial perturbation. If this ratio is smaller than one, the initial perturbation will decrease and the coexistence state is stable.  
In the general case, one has to consider the eigenvalues of the Jacobian matrix as in section~\ref{sec_app_jacobian}.

\section{Existence of stable coexistence region}
\label{sec_app_existence}
Our assumptions of symmetry and normalization allow us to write
\begin{equation}
 \omega_i(\gvec{\rho}^a)=\frac{\psi(\rho_i^a)}{\sum_{j=1}^{n_l} \psi(\rho_j^a)}
\end{equation}
where $\psi(\rho_i^a)$ is an arbitrary monotonically increasing function with $\psi(0)=0$, which is bounded on the interval $[0,1]$.
We have
\begin{equation}
  \left.\frac{\partial \omega_i}{\partial \rho_i}\right|_{\gvec{\rho}^{a*}} = \frac{\psi'(\rho_i^{a*})}{\psi(\rho_i^{a*})}  \frac{n_l-1}{n_l^2}             
  \end{equation}
which we can plug into Eq.~(6) to obtain
\begin{equation}
 \phi(\rho_i^{a*}) \equiv (1-\rho_i^{a*}) \frac{\psi'(\rho_i^{a*})}{\psi(\rho_i^{a*})} < 1 \eqdot
 \label{eqn_app_psi_condition}
\end{equation}
Since $\psi(0)=0$ and $\psi'(0) \neq \psi(0)$, the left-hand side of Eq.~\eqref{eqn_app_psi_condition} diverges 
for $\rho_i^{a*} \rightarrow 0$. Because $\psi$ is bounded, we have
\begin{equation}
\lim_{\rho_i^{a*} \rightarrow 1} \phi(\rho_i^{a*}) = 0 \eqdot
\end{equation}
Therefore, there is always a $\rho_i^{a*}$ (and so a value of $\lambda$) for which the inequality~\eqref{eqn_app_psi_condition} is fulfilled.


\end{document}